\documentclass[%
 reprint,
 amsmath,amssymb,
 aps,
 prl,
 longbibliography,
 lengthcheck,%
]{revtex4-1}

\usepackage{graphicx}
\usepackage{dcolumn}
\usepackage{bm}
\usepackage{hyperref}

\begin{document}

\preprint{APS/123-QED}

\title{African Dust Influence on Atlantic Hurricane Activity and \\ the Peculiar  Behaviour of Category 5 Hurricanes}

\author{ Victor M. Velasco Herrera }
\author{Jorge P\'erez-Peraza}%
\affiliation{Instituto de Geof\'isica, Universidad Nacional Aut\' onoma de M\'exico, Ciudad Universitaria, Coyoac\'an, 04510, M\'exico D.F., MEXICO}
\author{Graciela Velasco H.}
\affiliation{ CCADET, Universidad Nacional Aut\'onoma de M\'exico}
\author{Laura Luna Gonz\' alez}
\affiliation{Instituto de Geograf\' ia, Universidad Nacional Aut\'onoma de M\'exico}%

\begin{abstract}
We study the specific influence of African dust on each one of the categories 	 of Atlantic hurricanes. By applying wavelet analysis, we find a strong decadal modulation of African dust on Category $5$ hurricanes and an annual modulation on all other categories of hurricanes. We identify the formation of  Category $5$ hurricanes occurring mainly around  the decadal minimum  variation of African dust and in deep water areas of the Atlantic Ocean,  where hurricane eyes have the lowest pressure.  According to our	  results, future tropical cyclones will not evolve to Category 5 until the next decadal minimum that is, by the year $2015 \pm 2$.  

\end{abstract}

\pacs{Valid PACS appear here}
\maketitle


The variability of tropical cyclonic activity is ascribed to different factors, both natural and anthropogenic in origin  \cite{ref1,ref2}. 
The high activity since 1995 is not unusual in relation to previous periods; that's the reason why it is interpreted as a recovery of normal activity, instead of a response to the increase in  ocean surface temperatures   \cite{ref3}. Such is the case in the year 1998, when, according to satellite data, a maximum in temperature was reached, but only one Category $5$ hurricane  took place in that year. In contrast, four Category $5$ hurricanes occurred in $2005$ and two in $2007$.  During $2006, 2008$, and $2009$, no Category $5$ hurricanes took place, indicating that the situation is quite complex.

One of the more important factors associated with changes in the frequency of seasonal hurricanes is the variability of the local vertical tropospheric wind shear \cite{ref4}. In addition, a relation has been found between Atlantic hurricane activity and West African precipitation \cite{ref5,ref6} and with the SST anomalies associated with El Ni\~no \cite{ref7,ref8}. Recently, in order to contribute to the understanding of such variability several proposals have been offered related to cosmic rays and solar and geomagnetic activities \cite{ref9, ref10,ref11}; however, the main factor has not yet been well established.

The influence of African dust on Atlantic tropical cyclones has been proposed in \cite{ref12}. The dust from North Africa, which is transported over the Atlantic Ocean by easterly winds \cite{ref13,ref14}, contains high concentrations of small particles of the order of $10-100$ $g cm^{-3}$, which, in their way, change the properties of clouds and the evolution and development of precipitation \cite{ref15,ref16}. It also modulates    Caribbean storms \cite{ref2} and modifies the climate by absorbing and dispersing solar radiation. Together with other internal and external factors of the ocean-atmosphere system, such as winds and clouds  \cite{ref17}, African dust provokes a decrease in the ocean surface temperature, affecting the genesis of Atlantic tropical hurricanes \cite{ref18} by inhibiting their formation \cite{ref12}. 
\begin{figure}[b]
\includegraphics [width=0.48\textwidth]{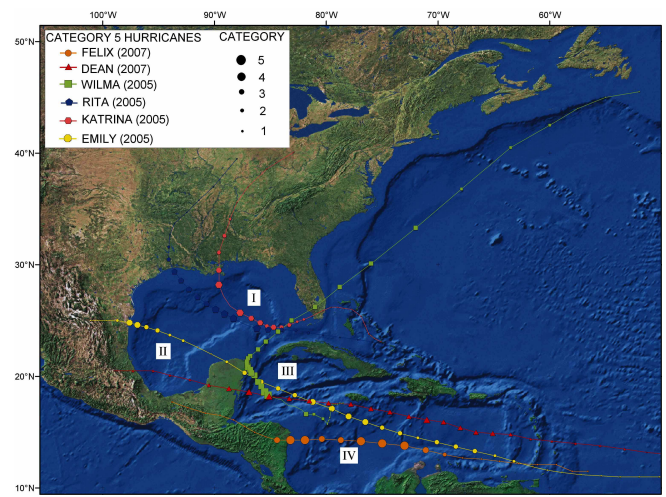}
\caption{\label{fig:epsart} 
Map of the Atlantic Ocean centers of cyclogenesis of Category $5$ hurricanes: I) the east coast of the United States, II) the Gulf of Mexico, III) the Caribbean Sea and IV) the Central American coast }
\end{figure}
Tons of dust from the deserts are transported by the winds over thousands of kilometers of the atmosphere. This dust interacts chemically with the clouds and radiation to modify the climate, while acting against global warming. The amount of dust increased considerably by the end of the $1960s$ and beginning of the $1970s$, when there was a severe drought in North Africa. 
Present climatic models that include African dust have shown that the variability of African dust is an important factor in predicting climatic change.

In spite of extensive studies that have been done on the influence of African dust on the whole planet, up to now a quantification has not been undertaken of the particular modulation of African dust on the evolution of  different hurricane categories. Here we evaluate the direct effect of dust on the frequency of each tropical cyclone category. 
\begin{figure}[b]
\includegraphics [width=0.5\textwidth]{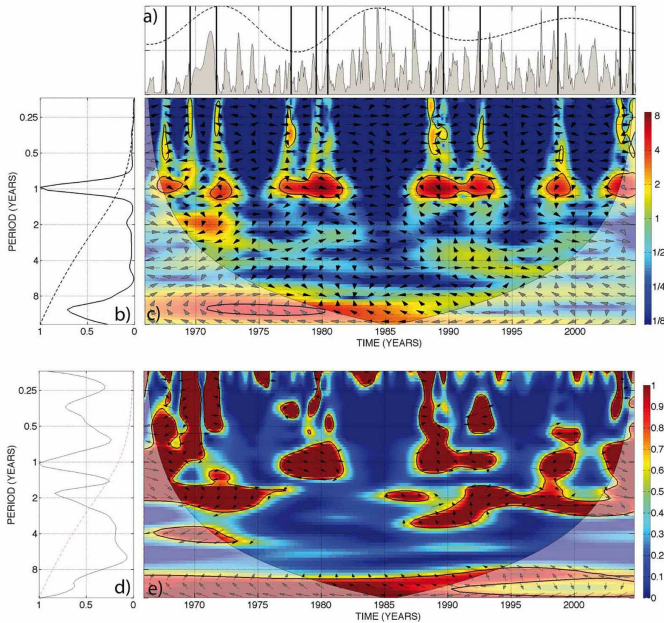}
\caption{\label{fig:epsart} 
(a) Time series of African dust (shaded area) - Category 5 Hurricanes (black bars) - decadal tendency of dust (dotted line). 
(b) Global spectrum of the cross wavelet transform (GXWT), 
(c) Cross Wavelet Transform (XWT) 
(d) Global Spectrum of Wavelet Transform Coherence-signal/noise (GWTCs/n),  (e) Wavelet Transform Coherence signal/noise (WTCs/n)}
\end{figure}
\begin{figure}[b]
\includegraphics [width=0.5\textwidth]{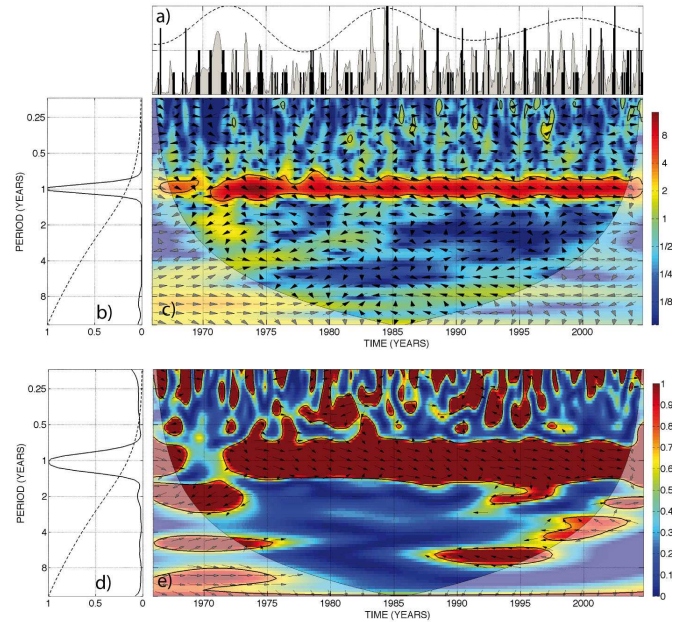}
\caption{\label{fig:epsart} 
(a) Time series of African dust (shaded area) - Tropical Storms (black bars) - decadal tendency of dust (dotted line). 
(b) GXWT
(c) XWT
(d) GWTC$_{s/n}$
(e) WTC$_{s/n)}$ }
\end{figure}
Data.-African dust  originates in the desert regions of the Sahara \cite{ref19,ref20}, the Republic of Chad \cite{ref21} and the uninhabited zones of Africa \cite{ref22}. Though there has existed satellite data on this dust since the $1980s$, in this work we use the only multi-decadal continuous in-situ dust monthly data available from Barbados from $1966$ to $2004$ \cite{ref23}.  The dust series is shown in gray area Figs. $2-7$.  A hurricane is a large mass of warm moist air with strong winds rotating in a spiral around a low-pressure area. When the water temperature reaches $26.5^o C$, mainly at latitudes of $10^o - 20^o$, the cyclogenesis of hurricanes occurs,  from the west coast of Africa [4] to I) the east coast of the United States, II) the Gulf of Mexico, III) the Caribbean Sea and IV) the Central American coast ( Fig. $1$). Their trajectories usually end at the mountains along the coast of the Gulf of Mexico and, to a lesser extent, at the mountains of the Caribbean islands. The well-known Saffir-Simpson scale gives a classification of hurricanes according to the intensity of their winds and the pressure in their eyes:

Depression: Sustained winds $\textless 63 km/h$ ($1008mb$ \textless $pressure$ \textless $1005 mb$). 
Tropical Storm: Sustained winds $63-118$  $km/h$ ($985mb$ \textless $pressure$ \textless $1004 mb$). 
Category One Hurricane: Sustained winds $119-153$ $km/h$ ($pressure$ \textgreater $980 mb$). 
Category Two Hurricane: Sustained winds $154-177$ $km/h$ ($965$ mb \textless $pressure$ \textless $980 mb$).
Category Three Hurricane: Sustained winds $178-209$ $km/h$ ($945 mb$ \textless $pressure$ \textless $965 mb$). 
Category Four Hurricane: Sustained winds $210-249$ $km/h$ ($920$ mb \textless $pressure$ \textless  $945 mb$). 
Category Five Hurricane: Sustained winds greater than $249 km/h$ ($pressures$ \textless  $920 mb$).

Fig. $1$ shows a map of the barometric distribution of the eyes of Category $ 5$ hurricanes illustrated  for six of these kind of hurricanes,  between $2006$ and $2007$ [Dean (2007), Felix (2007), Wilma (2006), Rita (2006), Katrina (2006) and Emily (2006)]. The black circles at the  right side  column indicate the evolution of each hurricane, from Category 1 to 5,  during their trajectory.
This data was taken from the National Weather Service  and transformed into a series of pulses with the technique PulseWidth Modulation (PWM) \cite{ref24} as: $n = $number of hurricanes, $0$ = no hurricane. This series is shown in panel ($a$) of Figs.  $2-7$.

Method. - In order to analyze local variations of power within a single non-stationary time series at multiple periodicities, such as the dust and hurricane series, we apply the wavelet (WT) using the Morlet wavelet \cite{ref25}. For analysis of the covariance of two time series $X$ and $Y$, we used the cross-wavelet $W_k ^{XY}$   (XWT), which is a measure of the common power between the two series \cite{ref26}.
The wavelet-squared transform coherence  $R_k ^2 (\psi)$  (WTC) is especially useful in highlighting the time and frequency intervals, when the two phenomena have a strong interaction \cite{ref27,ref28}. 
The WTC measures the degree of similarity between the input ($X$) and the system output ($Y$), as well as the consistency of the output signal ($X$) due to the input ($Y$) for each frequency component. 
When $R_k ^2 (\psi)=1$   , this indicates that all frequency components of the output signal ($Y$) correspond to the input ($X$). If   $R_k ^2 (\psi) \ll 1$  or  $R_k ^2 (\psi) \sim 0$ , then output $Y$ is not related to input $X$  because of the presence of noise, nonlinearities and time delays in the system.	
\begin{figure}[b]
\includegraphics [width=0.48\textwidth]{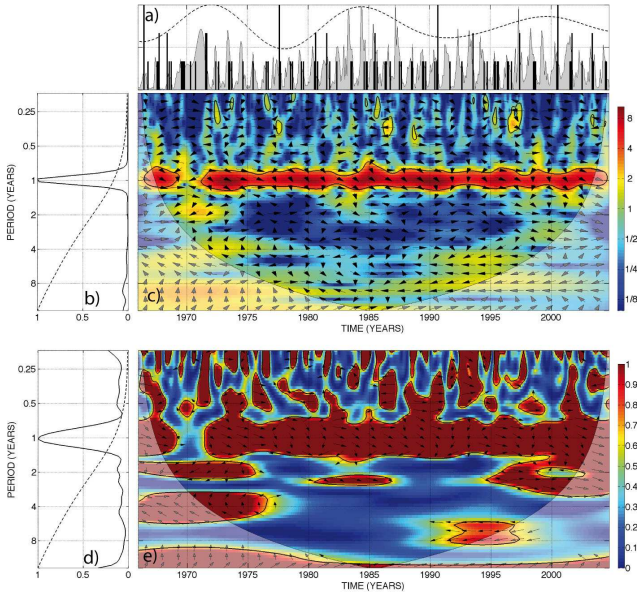}
\caption{\label{fig:epsart}
(a) Time series of African dust (shaded area) - Category 1 Hurricanes (black bars) - decadal tendency of dust (dotted line). 
(b) GXWT
(c) XWT
(d) GWTC$_{s/n}$
(e) WTC$_{s/n)}$ }
\end{figure}
The coherence of the system can be calculated through the relation signal/noise \cite{ref29}, such that we define the wavelet coherence signal-noise  $R_{s/n} ^2 (\psi)$  (WTCs/n)  as:	
\begin{eqnarray}
 \ {R_{s/n} ^2 (\psi)}=
\left(
\begin{array}{c}
 \  \frac{R_{k} ^2 (\psi)}{1- R_{k} ^2 (\psi)}  \ 
\end{array}\right)\
\end{eqnarray}
\begin{figure}[b]
\includegraphics [width=0.48\textwidth]{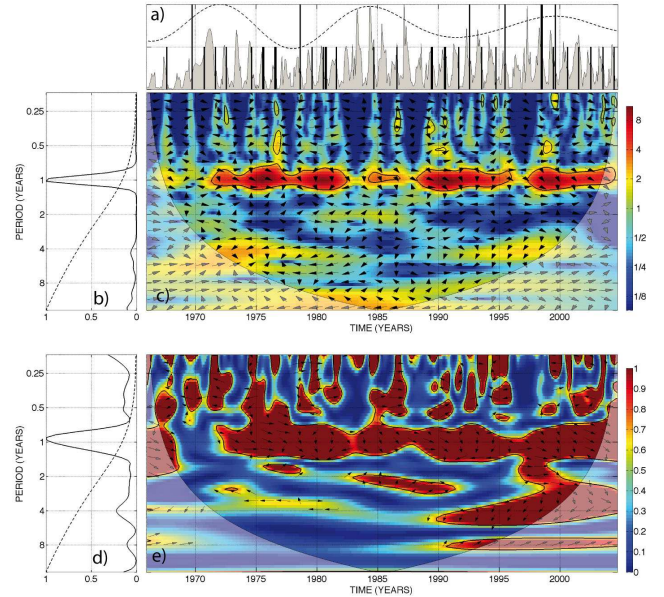}
\caption{\label{fig:epsart} 
(a) Time series of African dust (shaded area) - Category 2 Hurricanes (black bars) - decadal tendency of dust (dotted line). 
(b) GXWT
(c) XWT
(d) GWTC$_{s/n}$
(e) WTC$_{s/n)}$ }
\end{figure}		
\begin{figure}[b]
\includegraphics [width=0.48\textwidth]{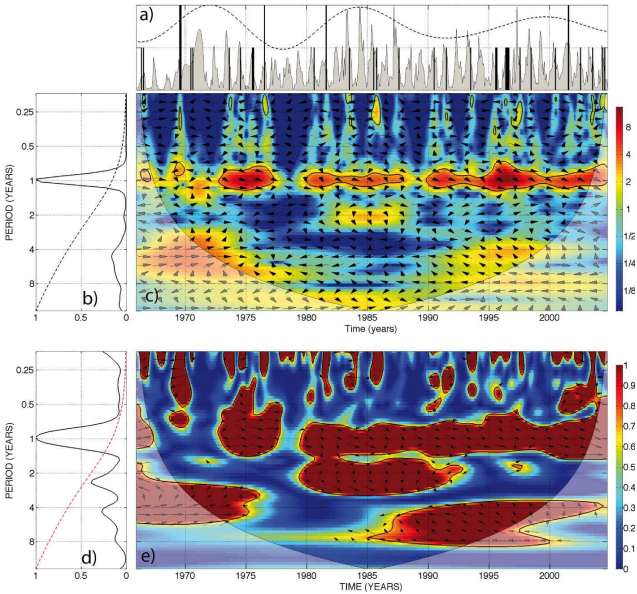}
\caption{\label{fig:epsart} 
(a) Time series of African dust (shaded area) - Category 3 Hurricanes (black bars) - decadal tendency of dust (dotted line). 
(b) GXWT
(c) XWT
(d) GWTC$_{s/n}$
(e) WTC$_{s/n)}$ }
\end{figure} 			
 The WTCs/n is what we are using in this study precisely because  it allows us to find linear and nonlinear relationships while verifying that the periodicities of cross-wavelet are not spurious in order to minimize the effects of noise.                                                                                                                                 	
If the XWT and the WTCs/n of two series are high enough, the arrows in the XWT and WTCs/n spectra show the phase between the phenomena. The arrows at $ 0^o$ (horizontal right) indicate that both phenomena are in phase, and the arrows at $180^o$ (horizontal left) indicate that they are in anti-phase. It is very important to point out that these two cases imply a linear relation between the considered phenomena; any other angles indicate that the two phenomena have a non-linear relation and a rather complex one at that. 
\begin{figure}[b]
\includegraphics [width=0.48\textwidth]{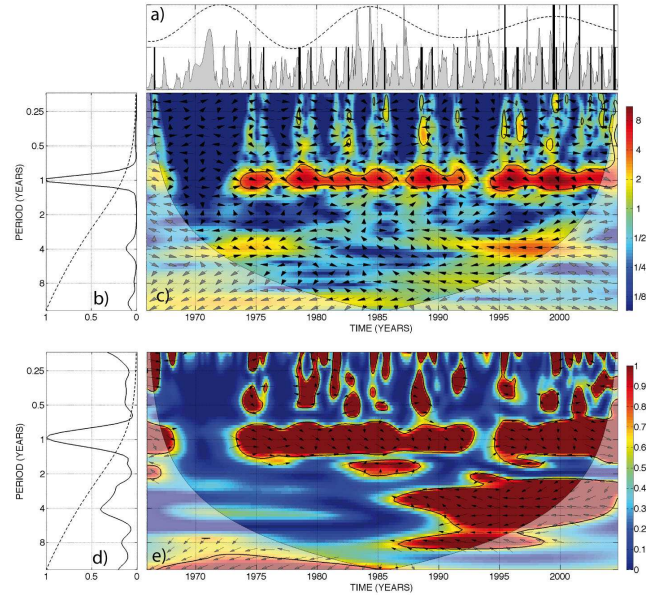}
\caption{\label{fig:epsart} 
(a) Time series of African dust (shaded area) - Category 4 Hurricanes (black bars) - decadal tendency of dust (dotted line). 
(b) GXWT
(c) XWT
(d) GWTC$_{s/n}$
(e) WTC$_{s/n)}$ }
\end{figure}
We also calculate the global spectra in the XWT (GXWT)  and WTCs/n (GWTCs/n), which is an average of the power of each periodicity.
 The significance level of the global spectra is indicated by the dashed curves (panels $c$) in the figures; they refer to the power of the red noise level at  $95$\% confidence, which increases with a decrease in frequency \cite{ref30}. 

Results. - Panel ($a$) in the following figures indicates the occurrence of hurricanes (black bars) according to the series of pulses calculated by the PWM technique: The dotted curve is the decadal behaviour of African dust, mentioned here after.

The GXWT (panel $b$ of Fig. $2$) between dust and Category $5$ hurricanes shows a very prominent annual periodicity with confidence higher than $95$\%. As can be seen in the XWT (panel $c$), this periodicity presents a high variability; it is not continuous and doesn't have the same intensity throughout the $1966-2004$ period, becoming more intense during he $1970s$ and at the beginning of the $1980s$, just when Category $5$ hurricanes were, on average, the  most intense. This annual periodicity is presumably related to the dust cycle in North Africa and to seasonal changes in atmospheric circulation \cite{ref31}. Additionally, there is a decadal periodicity ($11-13$ years), which is present throughout the entire time interval with an anti-correlation tendency (panel $c$). Its temporal tendency can be observed in panel ($a$) of Fig. 2 with a dotted line obtained by means of a Daubechies type modified wavelet filter This decadal periodicity is presumably related to the Atlantic trade wind variations and the dominant meridional mode of SST variability in the tropical Atlantic \cite{ref32}, as well as with solar activity \cite{ref7} and cosmic rays \cite{ref33}. The interaction of solar activity and cosmic rays with hurricanes is probably accomplished through the modulation of the Atlantic multidecadal oscillation \cite{ref9}.	
This decadal variation shows that Category $5$ hurricanes occur around the decadal minimum (panel $a$), because the local vertical wind shear ($V_z\geq 8m/s$) is unfavorable for the genesis of tropical cyclones \cite{ref7,ref34}. This would explain why from $2008$ up to the present, there have not been any Category $5$ hurricanes, since it was precisely during this time that  African dust in the atmosphere has been increasing. This would imply that if such a tendency continues, the next group of tropical cyclones will not evolve to category $5$ until the next decadal minimum of African dust occurs.

To confirm if the annual and decadal periodicities obtained with the cross-wavelet are intrinsically related to the modulation of African dust on Category $5$ hurricanes, we also obtained the modified wavelet coherence  (WTCs/n) and found, in addition to these two periodicities, two others, of $125$ days and $1.8$ years with a confidence level higher than $95$ \% (panels $d$ and $e$ of Fig. $1$). 

What we have is the existence of four areas of deep water in the Atlantic Ocean where the eye of the hurricane has the lowest pressure (\textless $920 mb$). This must be pointed out because it implies that, in addition to the required climatological conditions for the genesis of this kind of Category $5$ hurricanes to take place, the geography of the marine bottom also plays an important role, and that these hurricanes do not originate in hazardous places.

Fig. $3$ shows the GXWT (panel $b$) between dust and tropical storms where it can be seen that the most prominent periodicity is that of $1$ year. These annual variations do not have the same intensity throughout the period studied, as can be observed in the XWT (panel $c$) when the periodicity was low and parsimonious from the second half of the $1960s$ up to the first half of the $1970s$, due to an increase in precipitation in North Africa. At this point, the coherence becomes very intense due to the very severe droughts in West Africa, known as the Sahel drought, that began in the middle of the $1970s $ and lasted for several decades. The in-phase behaviour of this annual periodicity with a linear tendency seems to indicate that variability in dust has a quasi-immediate effect on the genesis and evolution of tropical storms. There are other periodicities lesser and greater than $1$ year but with  inconsistent patterns. This may be interpreted as that dust concentration and the evolution of tropical storms are the result of many external and internal factors occurring on different time scales. The XWT and WTCs/n show that the multiannual periodicities. Decadal periodicity is absent for  tropical storms. 

The global GXWT and WTCs/n (panels $b$ and $d$), as well as the cross-wavelet and coherence (panels $c$ and $e$) between dust and Category $1$ hurricanes (Fig. $4$) show a great deal of similarity with tropical storms, though with different intensities. These differences may be due to the increase of precipitation in North Africa, as indicated by the Sahel rainfall index [34]. Decadal periodicity is absent for Category $1$.

It can be observed in Fig. $5$ that the GXWT and GWTCs/n (panels $b$ and $d$), as well as the XWT and WTCs/n (panels $c$ and $e$) between dust and hurricanes for category $2$ show an annual periodicity that has a less continuous  time interval as compared to Category $1$ hurricanes (Fig.$ 4$) and tropical storms (Fig. $3$), becoming more intense from $1972-1982$ and $1988-2004$. Additionally, there exist less prominent periodicities of $3.5-5.5$ years that are associated with the El Ni\~no. Decadal periodicity is absent for Category $2$ hurricanes. 

The GXWT and GWTCs/n (panels $b$ and $d$), as well as the XWT and WTC (panels $ c$ and $e$) in Figs. $6$ and $7$ between dust and Category $3$ and $4$ hurricanes, respectively, show that the annual periodicity is also not continuous over the period studied. It can also be observed in panel ($e$) that the periodicity of  $3.5$ years for Category $3$ hurricanes becomes more intense during the interval $1980-1992$. For Category $4$ hurricanes the periodicities of $3.5-5.5$ years (panel $e$) are anti-correlated during $1988-2002$. Decadal periodicities for Category $3$ and $4$ hurricanes  are practically absent.
	
It should be emphasized that the observed correlations show not only a direct effect of African dust on hurricane activity but  also reflect an indirect relationship between  the wind, SST, AMO, the Modoki cycle, El Ni\~no, la Ni\~na, precipitation, solar activity and cosmic rays that to a greater or lesser degree modulate the evolution of Atlantic hurricanes.

Conclusions.-African dust influence on the genesis and evolution of Atlantic hurricanes varies in two main ways:
(a) annually for tropical storms and hurricanes of $1-4$ categories and\\
(b) decadally for Category $5$ hurricanes. 
These develop during the minimums of decadal cycles of  African dust.
In addition to peculiar climatological conditions, for future theoretical works and simulation with prognostic  goals, it must also be considered  the  geography of the Ocean bottom,
 where the four Atlantic deep-water regions, hurricane eyes have the lowest pressure.
This it is an important factor in the development and evolution of Category $5$ tropical cyclones, to be taken into account.
  
Finally, it is worth mentioning that if climatological tendencies continue as they have in recent decades, future hurricanes will not be able to develop into category 5 until the next decadal minimum that will begin in about $5 \pm 2$ years. 	\\

The authors wish to thank the Universidad Nacional Aut\'onoma de M\'exico (DGAPA-UNAM) for its support under grants IN119209-3, IN116705 and PE-105107.

\nocite{*}

\bibliography{IGF}

\end{document}